\documentclass{article}

\usepackage{arxiv}

\usepackage[utf8]{inputenc} 
\usepackage[T1]{fontenc}    
\usepackage{hyperref}       
\usepackage{url}            
\usepackage{booktabs}       
\usepackage{amsfonts}       
\usepackage{nicefrac}       
\usepackage{microtype}      
\usepackage{lipsum}
\usepackage{graphicx}
\usepackage{amsmath}
\usepackage{gensymb}
\graphicspath{ {./images/} }

\title{Weight Encode Reconstruction Network for Computed Tomography in a Semi-Case-Wise and Learning-Based Way}

\author{
 Hujie Pan \\
  School of Mechanical Engineering\\
  Shanghai Jiao Tong University\\
  Shanghai, China 200240 \\
  \texttt{phj1991@sjtu.edu.cn} \\
  \And
 Xuesong Li* \\
  School of Mechanical Engineering\\
  Shanghai Jiao Tong University\\
  Shanghai, China 200240 \\
  \texttt{xuesonl@sjtu.edu.cn} \\
  \And
 Min Xu* \\
  School of Mechanical Engineering\\
  Shanghai Jiao Tong University\\
  Shanghai, China 200240 \\
  \texttt{mxu@sjtu.edu.cn} \\
}

\begin{document}
\maketitle
\begin{abstract}
Classic algebraic reconstruction technology (ART) for computed tomography requires pre-determined weights of the voxels for projecting pixel values. However, such weight cannot be accurately obtained due to the limitation of the physical understanding and computation resources. In this study, we propose a semi-case-wise learning-based method named Weight Encode Reconstruction Network (WERNet) to tackle the issues mentioned above. The model is trained in a self-supervised manner without the label of a voxel set. It contains two branches, including the voxel weight encoder and the voxel attention part. Using gradient normalization, we are able to co-train the encoder and voxel set numerically stably. With WERNet, the reconstructed result was obtained with a cosine similarity greater than 0.999 with the ground truth. Moreover, the model shows the extraordinary capability of denoising comparing to the classic ART method. In the generalization test of the model, the encoder is transferable from a voxel set with complex structure to the unseen cases without the deduction of the accuracy. 
\end{abstract}


\section{Introduction}
Computed tomography (CT) with a limited number of views has found applications in understanding complex physics, such as turbulence and combustion in three-dimension \cite{cai2017tomographic, elsinga2006tomographic, floyd2011computed, li2014volumetric}. As a computational imaging approach, the most representative algorithms in reconstructing 3D scalar field from 2D images are the algebraic reconstruction technique (ART) and its derivations \cite{gordon1974tutorial, andersen1984simultaneous}. This category of deterministic algorithms requires a pre-calculated weight matrix to numerically correlate the pixel and voxel values. By iterating the ART method view-by-view, the difference between the original and projected pixel values is minimized to get the reconstructed voxel set. A critical issue of the ART-based algorithm is that the weight matrices require considerable sizes of RAM of the computer, which may easily go to tens of GBs and exceed practical hardware limitations \cite{li2014volumetric}. A second critical issue is that the limited computation resources, together with the inadequate domain knowledge, lead to the mismatch of the assumptions of weight matrix calculation and real-world physics, which eventually results in the inaccuracy of the reconstructed 3D voxels \cite{trull2017point,zhang2007gaussian, li2015capabilities}. Moreover, the deterministic nature of the ART-based algorithm makes the reconstruction vulnerable to noise and systematic error. 

Deep learning, as its extraordinary capability of feature abstracting and nonlinear fitting \cite{lecun2015deep}, can potentially deal with the issues mentioned above. However, a typical supervised learning method calls for a large amount of data with properly assigned ground truth, which cannot be easily fulfilled under many conditions. Some attempts have been made to adopt deep learning algorithms in some reconstruction tasks like the sliced absorption spectroscopy and chemiluminescence reconstruction \cite{huang2020limited, huang2018reconstruction}. However, such end-to-end methods (using the images as the input and voxels as the outputs) potentially deduct the generalization of the model. It limits the scope of the application of the model and makes it almost not possible to be transferred to unseen cases or those with different numerical setups. \cite{wurfl2018deep} tried to increase the generalization of the model with a more built-in projection manner, which helps the model learn a more analytical representation from the projections. However, it is still limited by the end-to-end form that requires well-labeled data and fixed setup. 

To tackle the issues mentioned above, we propose a semi-case-wise learning-based method for limited view computed tomography, named Weight Encode Reconstruction Network (WERNet). Unlike the end-to-end methods that compute the 3D voxels directly by 1-stage inference, we build a brand-new semi-case-wise architecture that learns the voxel value in the training process. In this model, different from typical deep learning models, the sought voxel values are co-trained as parameters in the neural network together with the weight matrix (namely the voxel weight encoder)in an unsupervised manner. Gradient normalization was used to distinguish the two types of parameters and stabilize the training process. We then successfully realize the co-training of voxel values and voxel weight encoder. 

The WERNet algorithm shares some similarities with the NeRF algorithm for 3D object surface reconstruction \cite{mildenhall2020nerf} but they are intrinsically different. The reconstruction 3D light field in the NeRF model was implicitly incorporated in the neural network, while the 3D flame light field (intensity information) is explicitly and quantitatively set as the parameters in the WERNet algorithm. The neural network consists of hybrid parameters, including explicit flame intensity distribution and implicitly expressed weight encoders that stand for the physics of light propagation and imaging. It is worth mentioning that the flame intensity parameters in the neural network are physically meaningful. These parameters are obtained as the neural network is trained, which sheds light on solving similar physical applications with deep learning algorithms besides using traditional end-to-end algorithms. 

By adopting the novel WERNet method, the voxel set is successfully learned in a single case with the cosine similarity of the reconstructed voxel set to the reference greater than 0.999. Moreover, the RAM requirement is dramatically reduced since the tremendous weight matrix is no longer needed to be stored. Without the deterministic nature, WERNet has a much better capability of denoising than that of ART-based methods, especially the background, which yields a sharper and more accurate reconstructed voxel set. Finally, we freeze the pre-trained weight encoder and transfer it to unseen cases to evaluate the generalization of the method. It turns out that our method is transferable under different conditions, and it is able to achieve comparable performance in the reconstruction with faster convergence, benefiting from pre-trained, and optimizable weight encoder.

\section{Method}
The WERNet model is a two-step algorithm including data preparation and model training. The data preparation section introduces how we set the case and preprocess the data to feed WERNet for training. 

\subsection{Data Preparation}
To generate the data with the reference for training and test, we adopt the classic computational photographic method to project the reference voxel set to different angles of views with pre-determined weight matrix. More specifically, in this work we used the pinhole model for ray-calculating purpose. It will not affect the generality of our method since our method requires no information of the reference voxel set and its numerical correlation to the projections as the input or initialization. In the current model, each pixel corresponds to one ray only that across the pinhole, while it is readily to establish the correlation between a pixel and multiple rays in the WERNet model.

\subsubsection{Impacting Voxel Search}

Unlike a typical end-to-end method that builds the correlation between the voxel value and pixel values directly using a deep neural network, the WERNet model adopts a sequence of impacting voxels of a pixel as the inputs of the algorithm, and the pixel values as the outputs. The weight matrix (representing the imaging physics) and the 3D light field (flame intensity) were considered as the parameters trained in the model. 

\begin{figure}[htbp]
\centering
\includegraphics[width=3in]{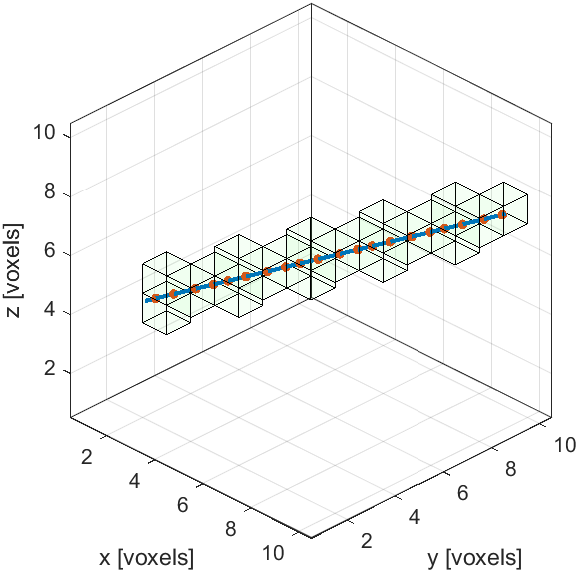}
\caption{Schematic of voxel search and seg points. The blue line is the pseudo ray and the green cubes are the impacting voxel. The seg points are represented with red points. }
\label{fig:vsearch}
\end{figure}

As seen in Fig. \ref{fig:vsearch}, we employ the ray-tracing method to obtain the impacting voxels. The blue line is the reversing ray coming from a pixel through the aperture. When the ray passes through the voxel set, a certain sequence of impacting voxels can be extracted as labeled with green faces in Fig. \ref{fig:vsearch}. Each impacting voxel and the reversing ray intersect at a segment. We record the coordinates of the center (seen as red points in Fig. \ref{fig:vsearch}) of the segment for further feature abstracting. With the above procedure, we gather the information of the impacting voxel sequence with the latent information of the ray trace and view angle of the camera. It is worth noting that in this arrangement, the ray can be a complex trace, i.e., impacted by refraction, which makes the WERNet model compatible with more challenging reconstruction tasks than traditional methods.

\subsubsection{Sample Data Generation}
For the impacting voxel sets that correspond to different pixels, the lengths of them can vary. To efficiently feed the data to the WERNet model, we preprocess the data by extending the shorter voxel sequence with a zero tensor and make all the lengths of sequences consistent. For instance, 

\begin{equation}
input = \begin{bmatrix}
\textbf{I}^{3 \times n} & \textbf{O}^{3 \times (N-n)} \\
\textbf{P}^{3 \times n} & \textbf{O}^{3 \times (N-n)}
\end{bmatrix}
\label{inputtensor}
\end{equation}

in which $\textbf{I}^{3 \times n}$ is the indices of the voxels, $\textbf{P}^{3 \times n}$ is the coordinates of the seg points of the voxels and $\textbf{O}^{3 \times (N-n)}$ is the zero tensor for the extension. So the concatenated tensor is of the dimension 3-by-N. By doing so, we normalize all inputs to $6 \times N$ for training the WERNet model.

\subsection{WERNet}
\subsubsection{Architecture}

\begin{figure}[htbp]
\includegraphics[width=\linewidth]{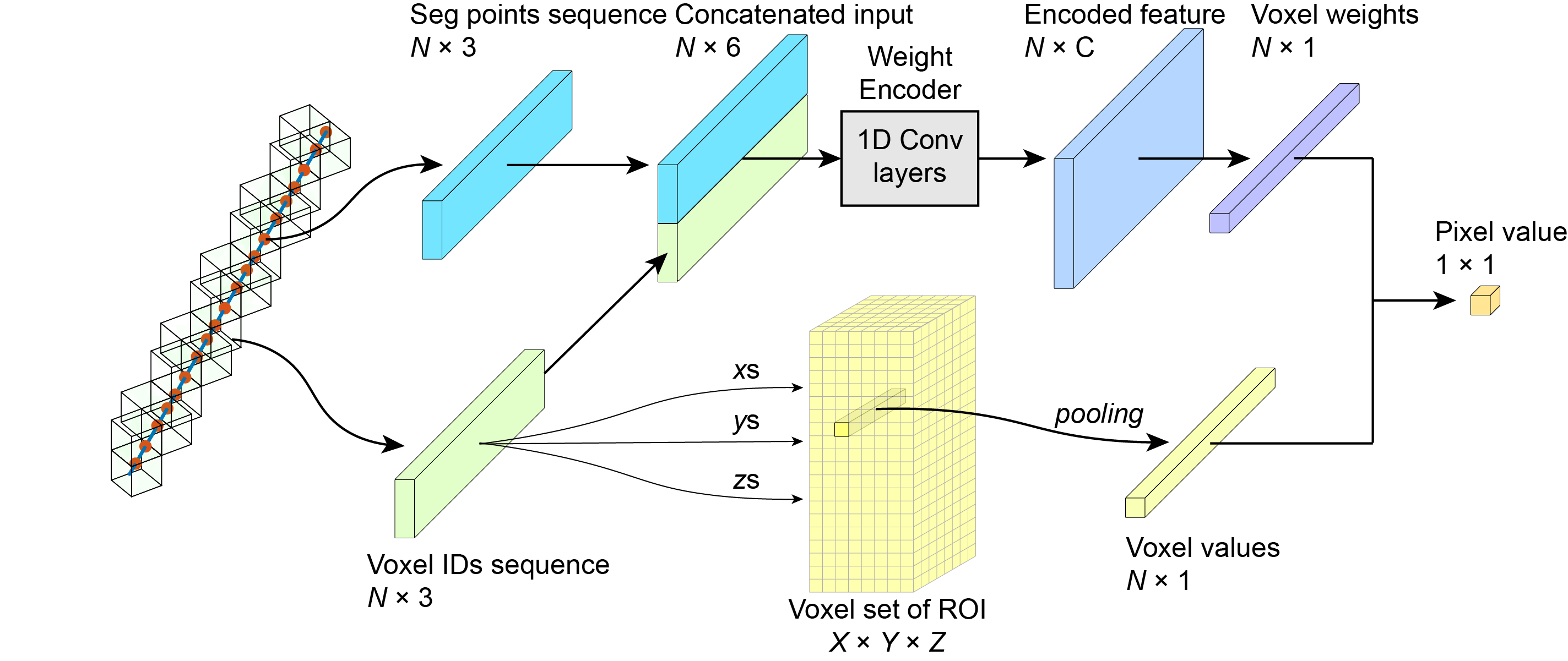}
\caption{Architecture of the network. The cuboids in the figure represent the tensors of different dimensions and the arrows represent the direction of the data flow.}
\label{fig:archi}
\end{figure}

The architecture of the WERNet model is organized into two sections. The first is the weight encoder section, as seen in the upper region of Fig. \ref{fig:archi}. For each block the ray passes through, we use a 1-D convolution layer to abstract the features of the inputs and Leaky ReLU for non-linearity. The kernel size of the convolution layer is 3, and the output channel 32. We also set the padding and stride both 1 to make the length of output the same with the input. These parameters have been tested and optimized. After outputting the 1-D tensor of the same dimension with the voxel value, we square the result to make sure that the weights are always non-negative. 

The other section is for voxel attention and pooling, shown in the lower portion of Fig. \ref{fig:archi}. We determine the impacted voxels using the indices obtained from the target ray information. Then we pool the voxels by extracting these voxels from the voxel set for further computation. It is worth noting that we extend the sequence of voxel values with a zero tensor to make its dimension consistent with the voxel weights. 

Finally, we conduct an element-wise multiplication of voxel weights and voxel values and then sum them up to obtain the predicted pixel value of the projection.

\subsubsection{Gradient Normalization}
For the section of calculating the pixel value mentioned previously, we modified the gradients of the voxel value in the following form: 

\begin{equation}
    \begin{split}
    g_{i,output}^{vi} &= g_{i,output}^{vi} \cdot w_{i} / \text{norm}(\textbf{w}) \\
    g_{i,output}^{wi} &= g_{i,output}^{wi} \cdot v_{i}    
    \end{split}
    \label{gradientnorm}
\end{equation}

in which $g_{i,output}$ is the output gradients, $g_{i,input}$ is the input gradient from the last backpropagation step. $w_i$ or $v_i$ is the multiplication coefficient of the $i$th voxel value or weight. 

For the voxel weight, we maintain the original gradient form to keep the training manner for the weight encoder. As for the voxel value, we add norm(\textbf{w}) to normalize the original gradient to stabilize the training process. 

The modified gradient method proposed offers different optimization schemes for voxel values and the weight encoder, which plays an essential role in remedying the model. More detailed results of the comparison between modified and unmodified gradient form can be found in the results section. 

\subsubsection{Learning Algorithm}

We compare Adam and SGD optimizers with different learning rates and adopt Adam in the model. For the voxel value, since we use the normalized gradients, we set the learning rate of 0.01 to accelerate the convergence. As for the other parameters of the model, we set a learning rate of 0.0005 to avoid numerical instability. Moreover, we set a decay rate of 0.5 for the learning rate for every five epochs to finetune the model. 

The batch size of the training process is 32. Notice that each sample contains 100 pairs of input sequences and targets, which indicates an equivalent batch size of 3200. With the hyperparameters of the learning algorithm set as above, we train our model for 80 epochs to get the final result. 

\section{Results}
\subsection{Numerical Setup and Metrics}
\subsubsection{Numerical Setup}

\begin{figure}[htbp]
\includegraphics[width=\linewidth]{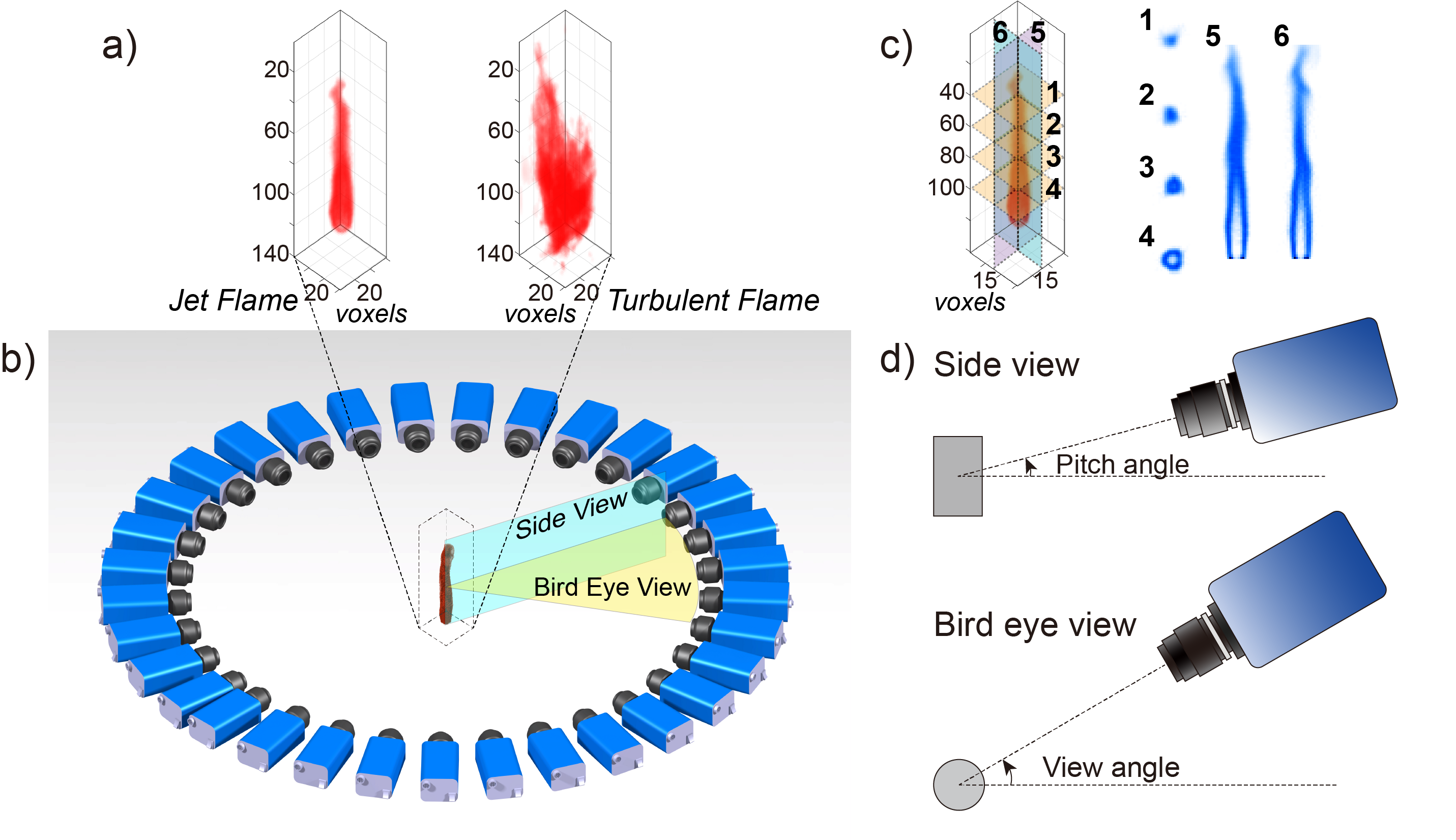}
\caption{Experimental setup of virtual cameras and reference voxel sets, a) original voxel sets; b) layout of the virtual cameras (views); c) explanation of the cross sections figure; d) explanations of pitch angle and view angle.}
\label{fig:setup}
\end{figure}

To evaluate the WERNet model, we conducted a series of numerical experiments to explore its performance. The domain of interest (where the flame or flow is set) was discretized into a $30\times140\times30$ mesh with a size of 0.5 mm for reconstruction. The view angles (defined as seen in Fig. 1d) were set with a resolution of $11\degree$ around the object starting from the origin $0\degree$. The distance from the camera to the object was set around 5800 mm. The pitch angles of the cameras (as defined in Fig. \ref{fig:setup}d) were set either $0\degree$, $15\degree$ or $-15\degree$. 

We verify our method on two simulated flames in the form of voxel set with different geometries and complexities, as reflected in Fig. \ref{fig:setup}a. One is a jet flame with a simpler and more symmetric geometry, and the other one is a turbulent flame that is more disordered and asymmetric. Projections on the camera were simulated from the ground truth. The WERNet used the projections to reconstruct the flame just like other algorithms, and the reconstruction is compared with the ground truth in justifying the capacity of the WERNet method.  

To quantify the discrepancy of the reconstructions, we calculated the cosine similarity of the predicted voxel set to the original one(ground truth) by:

\begin{equation}
    S_C = \frac{\textbf{v}_o \cdot \textbf{v}_r}{\left \| \textbf{v}_o \right \| \left \| \textbf{v}_r \right \|}
\end{equation}

in which \textbf{v} is the vector of flattened voxel set. The subscripts $o$ and $r$ represent original and reconstructed voxel sets, respectively. The difference between the reconstructed voxel set and the reference can be represented using cosine distance, which is defined as $D_C = 1-S_C$. 

\subsection{Algorithm Verification and Parametric Study of View Layouts}

We varied the layout of the virtual cameras to capture the multi-view images with different arrangements. For the jet flame voxel set, we conducted three different numerical experiments to test the performance of our model with the same number of views but different poses and distances of the virtual cameras. The first case is with 33 views that uniformly arranged around the object at a constant distance of 5800 mm from the object. The second case is almost the same as the first one but with a preset pitch angle of  $15\degree$ or $-15\degree$. The third one adopted random distance on the basis of case 2 for higher complexity. 

\begin{figure}[htbp]
\includegraphics[width=\linewidth]{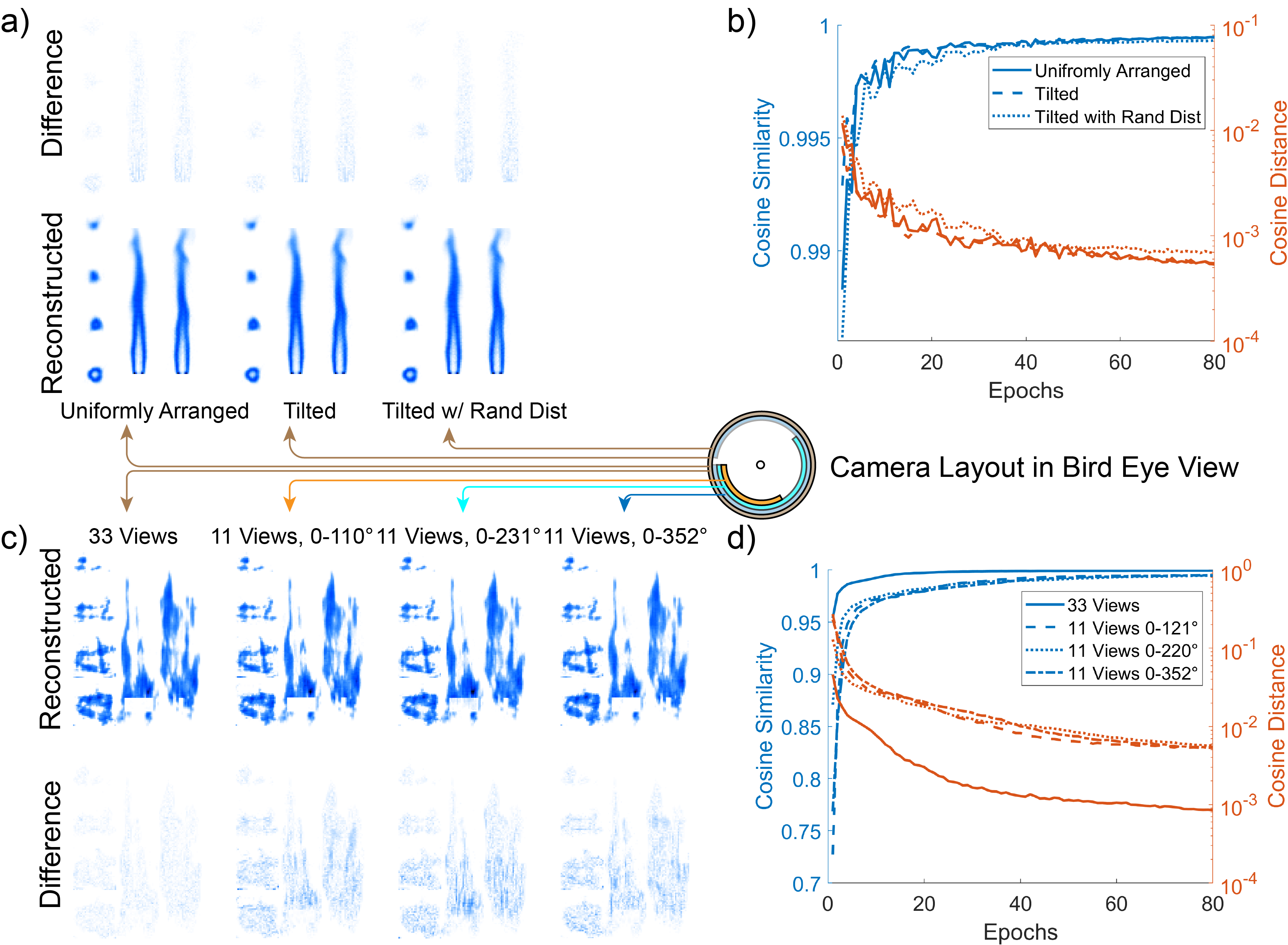}
\caption{Model performance under different layout of the views. The rings of different colors represent the camera layout in the bird eye view. a) and b) are the cross sections of the reconstructed voxel sets of jet flame and turbulent flame, respectively. c) and d) are the cosine similarities of reconstructed result and the reference. }
\label{fig:geperform}
\end{figure}

The cross-sections of the reconstructed flame and their difference with the ground truth are shown in Fig. \ref{fig:geperform}a. For all the three layouts, the WERNet algorithm successfully reconstructed the ground truth of the jet flame. The reconstruction performance appears to be insensitive to pitch angles and the object distances, demonstrating the capacity of the WERNet algorithm. The cosine distances plotted in Fig. \ref{fig:geperform}b goes below $10^{-3}$ as the epochs increase to 80, which quantitatively demonstrates the high accuracy of the reconstructed results. 

We further examine the WERNet algorithm with the turbulent flame phantom. The influence of view numbers is explored, as well as their layouts on the reconstruction results. The case with 33 views is designed with a resolution of $11\degree$ for view angles, $15\degree$ and $-15\degree$ alternately for pitch angles, and random distances ranging from 5500 mm to 6500 mm. As for the cases with 11 views, we test three different view angle combinations: (1) $0\degree$ to $121\degree$ with a resolution of $11\degree$; (2) $0\degree$ to $220\degree$ with a resolution of $22\degree$; (3) $0\degree$ to $352\degree$ with a resolution of $33\degree$. As shown in Fig. \ref{fig:geperform}c and \ref{fig:geperform}d, the WERNet accuracy is sensitive to the view number used, but not quite relevant to the view arrangements.

\subsection{Ablation Study}
\subsubsection{Influence of Batch Normalization in the Encoder}

As mentioned in the Method section, the impacting voxel sequences are used as the inputs of the WERNet algorithm. The lengths of impacting voxel sequences of different pixels depend on the actual number of voxels that the pseudo ray passes through. To make the lengths of each voxel sequence consistent, we extended the shorter sequences with a 0-value tensor. For handling the heterogeneous inputs, we tested three methods to keep the extended part always 0 in the inference so that no gradient will be backpropagated during the training process. Method 1 deactivates the biases of the 1-D convolutional layers to make the convolution result of the extended part always 0. Method 2 activates the bias while using a mask to manually set the encoded voxel weights of the extended part 0. Similarly, Method 3 deactivates the bias while adding batch normalization (BN) layers to rescale and shift the features. 

\begin{figure}[htbp]
\centering
\includegraphics[width=5in]{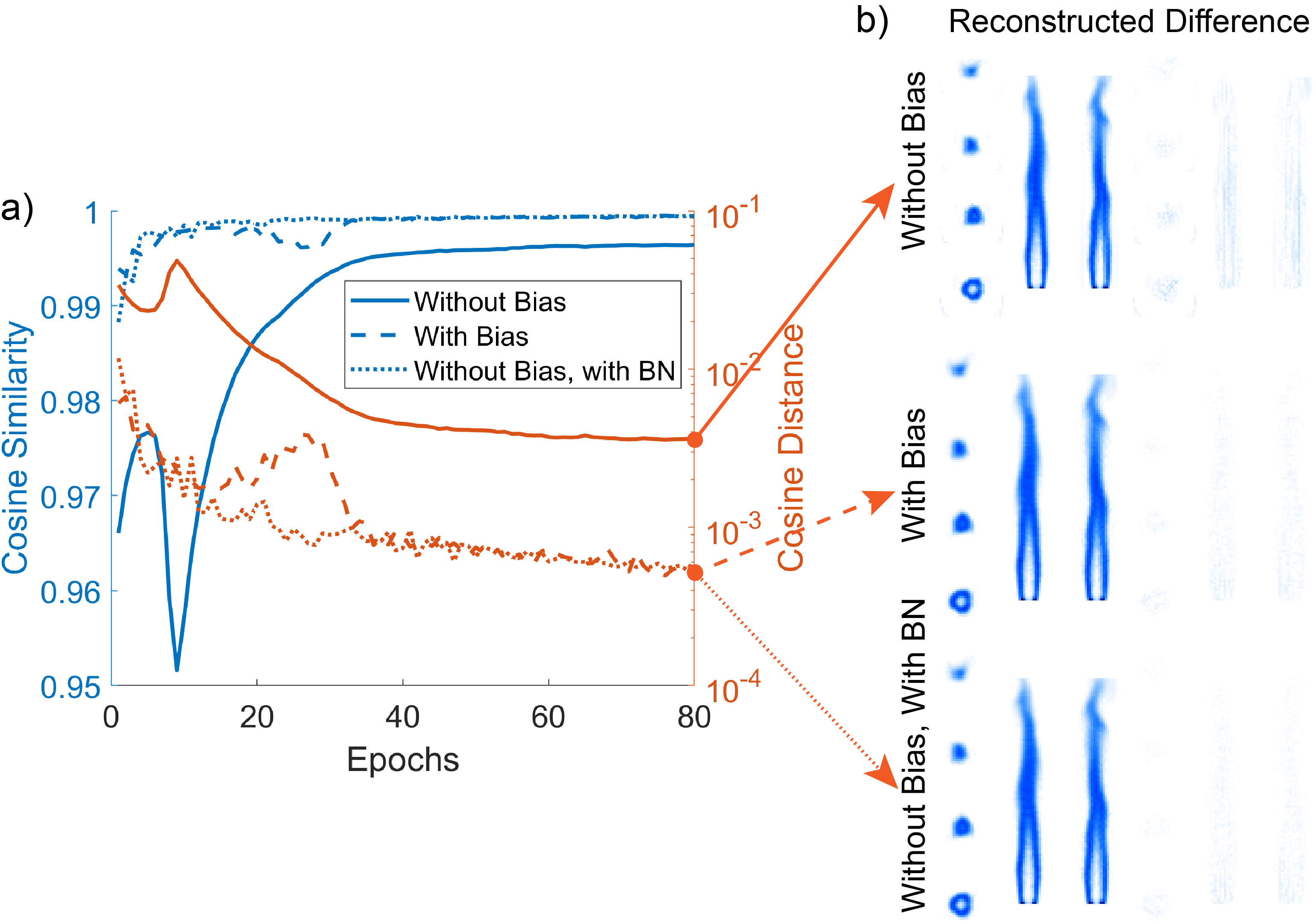}
\caption{Comparison of different methods adopted in the weight encoder. a) the cosine similarity of the reconstructed results and the reference. b) the final reconstructed results and their absolute difference from the reference.}
\label{fig:encoder}
\end{figure}

As we can see in the similarity shown in Fig. \ref{fig:encoder}a, the model without biases or BN layers has the worst performance due to the deducted capability of nonlinear fitting. Method 2 and 3 achieve a similar level of performance after 40 epochs (with the similarities finally surpass 0.999), which indicates an ideal reconstruction result of the voxel set. The equivalent large batch size makes it a good representative of the whole dataset and empowers the BN layer to stabilize the training process by rescaling the intermediate features and reducing the internal covariate shifts \cite{ioffe2015batch}. In that case, the adoption of the BN layer yields a more robust training process than Method 2. 

\subsubsection{Influence of Gradient Normalization}

In the WERNet network, two types of parameters with completely different physical implications were incorporated. The first category is the weight encoder, which represents the physics of light propagation (similar to the weight matrix in traditional CT algorithms but able to represent more complex non-linear physics). The second category, as we discussed, is the scalar flame intensity itself, which represents the 3D distribution of the flame intensity numerically. Therefore, for this specific application using WERNet, two different gradient flows are needed for co-training. 

\begin{figure}[htbp]
\includegraphics[width=6in]{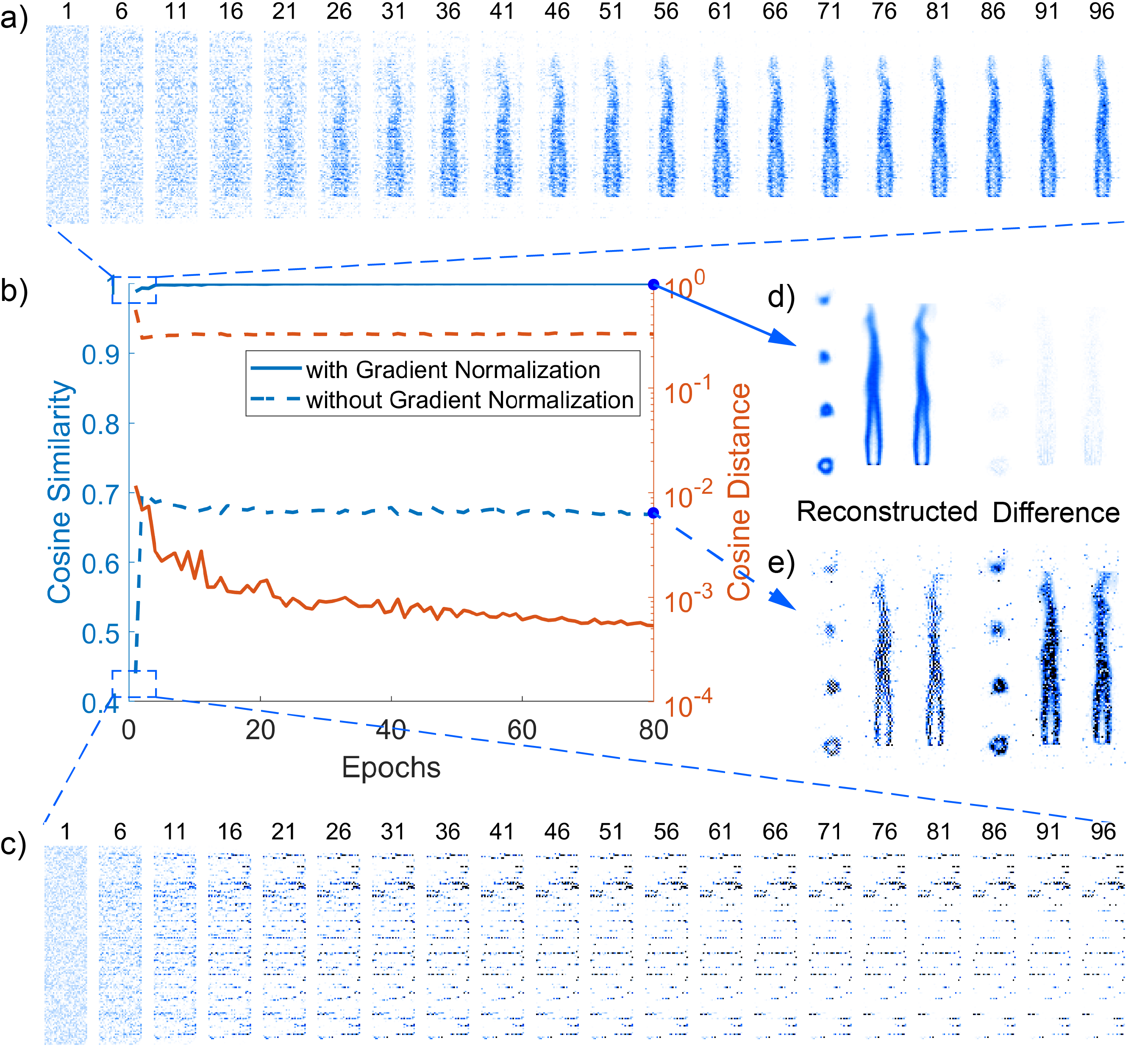}
\caption{Comparison of original gradient form and normalized gradient backpropagation. a) and c): first 96 iterations in the first epoch with and without gradient normalization; b) the similarity of reconstructed voxel set and the reference; d) and e) cross sections of reconstructed voxel sets and their difference from the reference. }
\label{fig:artback}
\end{figure}

As mentioned in the Method section, we normalize the gradient for voxel values while keeping the original gradient form for the encoder. The image sequence in Fig. \ref{fig:artback}a and c show the comparison of the results between the methods with and without gradient normalization. With the original gradient form, the voxel value is trapped after the first ten iterations, which is determined by the first fed batches and the initialization of the voxel weights. Actually, due to the simple multiplication operation of voxel weights and voxel value, the updating rate of the voxel value is sensitive to the inferred voxel weights from the encoder. Therefore, the discrepancy of the encoded voxel weight in the early stage can cause the voxel values significantly deviate from the ground truth. As a result, the original gradient form would cause a slower convergence, and the results might be trapped to local minima.  

In contrast, by normalizing the gradient for the training of voxel values, the updating rate of the voxel values is flattened. Therefore, the whole training process is stabilized. As a result, the voxel set with gradient normalization shows extraordinary convergence speed comparing to that with the original gradient form. 

As for the similarity of the reconstructed results to the reference shown in Fig. 4b, the method with gradient normalization significantly outperforms the method using the original gradient form. As seen in Fig. \ref{fig:artback}b and \ref{fig:artback}e, the method using the original gradient fail to reconstruct the jet flame in 80 epochs. The trend of the similarity curve shows a low possibility for the method to successfully escape the local minima. On the contrary, the WERNet method that adopts gradient normalization successfully learns the voxel values and voxel weight encoder. 

\subsection{Capability of Denoising}

To evaluate the capability of the WERNet algorithm in denoising, we add a Gaussian distributed noise to the projection with a magnitude of around 10\% of the maximum of the projected intensities, which is a relatively harsh noise level for reconstruction purposes. 

Generally, the WERNet method outperforms the classic ART method in reconstructing the 3D voxel set with noisy projections in both jet and turbulent flame cases. For the jet flame, the WERNet method produces the result with minor deterioration in accuracy comparing to the case with clean projections. While the classic ART method shows a  high sensibility to the noise added, due to its deterministic nature and the flaws of the reconstructed result arise not only in the non-zero region (flame region) but also the zero-value background in the ground truth. 

\begin{figure}[htbp]
\includegraphics[width=\linewidth]{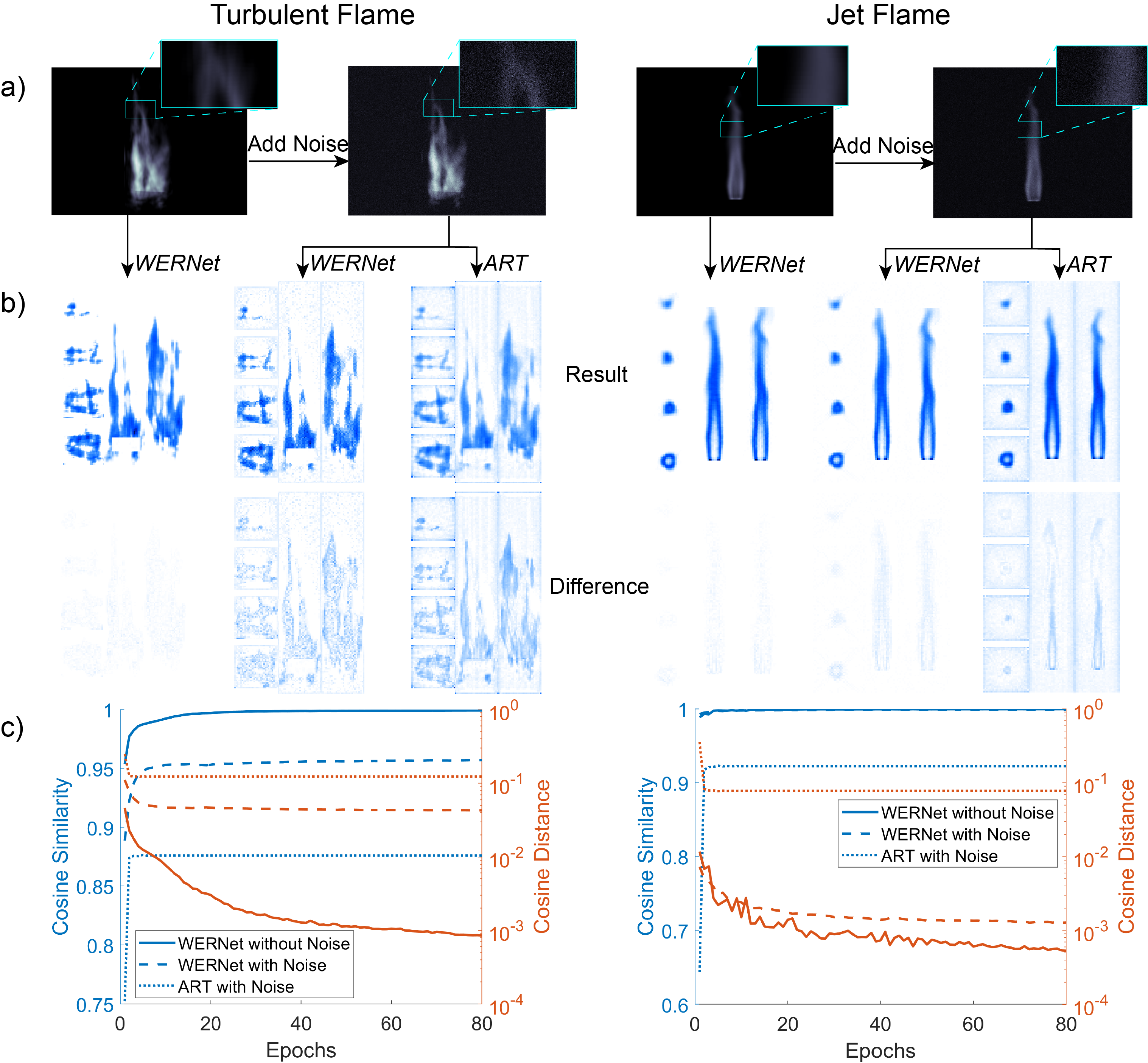}
\caption{The comparison of the capability of denoising between ART method and WERNet. a) clean noisy projections. b) the cross sections of reconstructed results and their difference from the reference. c) the cosine similarities of the reconstructed results and the reference at different epochs. }
\label{fig:denoise}
\end{figure}

In comparison to the jet flame cases, the voxel values of the turbulent flame are more chaotic and less smoothly distributed, which means the maximum-intensity-based noise is of higher relative magnitudes in turbulent flame cases. The WERNet method reconstructs the flame with a similarity greater than 0.95 to the ground truth with noise added, while the ART method achieves a performance of around 0.87 only. For more details in the patterns of the voxel sets, as seen in Fig. \ref{fig:denoise}b, the WERNet method reconstructs the turbulent flame with higher contrast and less noise in the background comparing to the blurred, noisy result from the ART method. 

With more non-zero projected intensities, i.e., valid data for training the voxel weight encoder (otherwise the gradient would be 0), the reconstructed result converges faster in noisy cases than that without noise, as seen in Fig. \ref{fig:denoise}c. In the case of turbulent flame, due to the noisier projection comparing to that of jet flame, the similarity converges in less than 10 epochs and then changes minorly as the number of epochs increases.

\subsection{Model Transfer}

As seen in the above sections, the WERNet method successfully learns the intrinsic nature of light propagations and imaging while accurately reconstructing the voxel set as the model is being trained. The parameters in the network, as a reminder, are flame intensities and weight encoders, respectively. The flame-related parameters are case-dependent and not transferrable to other cases. While with no change of the physics in producing the projection of the reconstructing region, the weighted encoder is intuitively transferable to different cases. 

\begin{figure}[htbp]
\includegraphics[width=\linewidth]{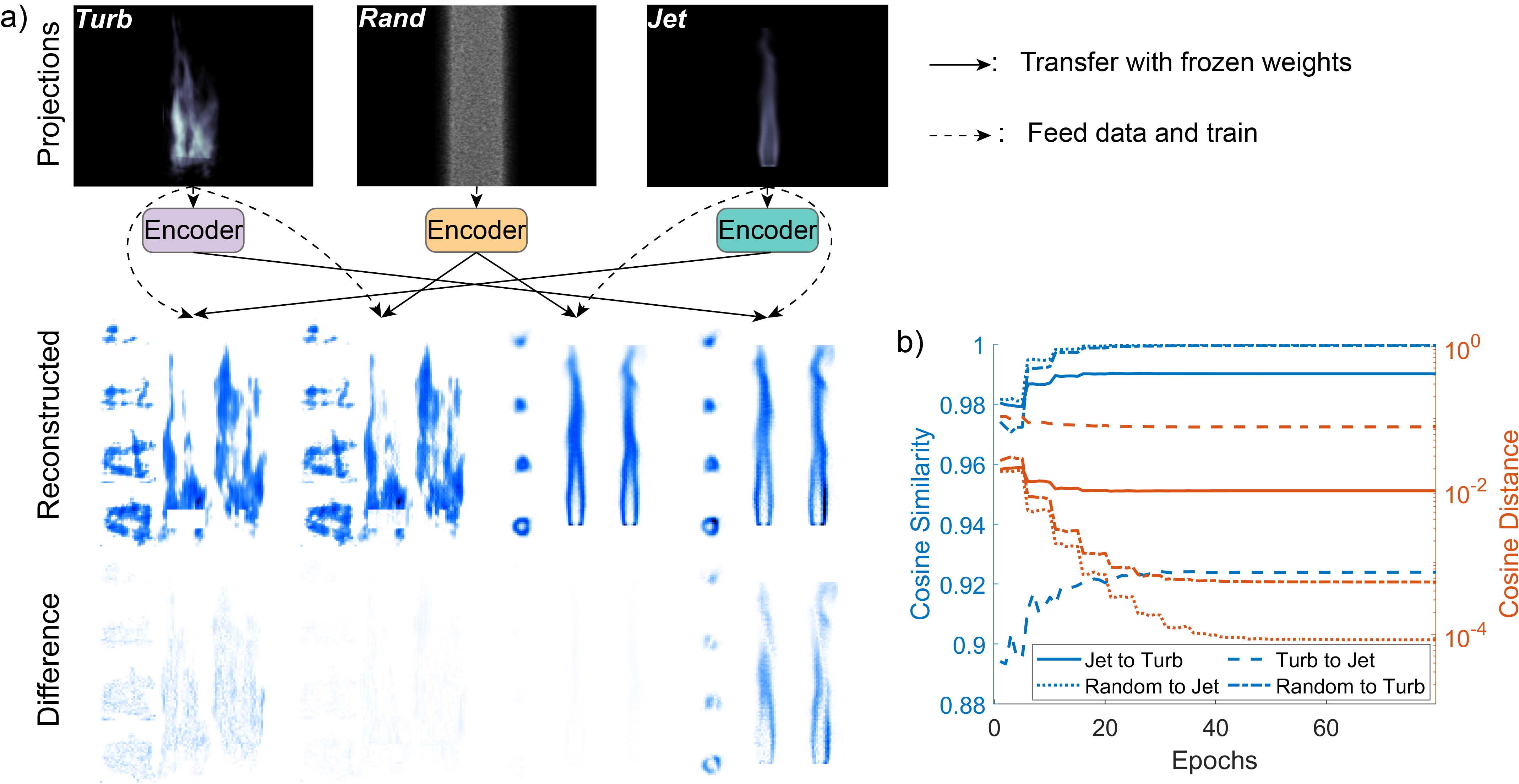}
\caption{Test of model transfer of WERNet. a) images in the first row are the projections of three different cases named turbulent flame, randomized voxel and jet flame. The second and third rows contain the reconstructed result and their difference from the reference. b) the cosine similarity of the results of different transferring cases and the reference at different epochs.}
\label{fig:transfer}
\end{figure}

To examine this assumption, we freeze the voxel weight encoders obtained previously and only modify the value of the voxel set, which is randomly initialized. With the voxel weight encoder trained in the jet flame case, the turbulent flame is successfully reconstructed with a final similarity close to 0.99 to the reference after 80 epochs. While when we transferred the weight encoders from the turbulent flame cases to the jet flame, the final similarity stops increasing at 0.92 as seen in Fig. \ref{fig:transfer}b. Besides, neither of the results reaches the similarity that achieved in the case with trainable voxel weight encoder (>0.999). 

In the experiments we conducted previously, it should be noted that zero value pixels or voxels have a very limited contribution to the model training process. Therefore, for the transferred weight encoders to function, non-zero pixel and voxel values should be used during the training. When transferring the encoder of the non-homogeneous biased turbulent flame, the reconstructed jet flame become more asymmetric comparing to the reference. However, since the jet flame is relatively rotationally symmetric which agrees with the isotropic manner of voxel weight derivation in our study, the reconstructed turbulent flame using jet flame’s encoder reaches a higher similarity and the difference of it from the reference evenly distributed in the non-zero voxel region.

To further examine the hypothesis, we built a randomized homogeneous voxel set, as seen in Fig. \ref{fig:transfer}a, which filled the volumetric domain of interest (where the flames are expected to occur) to train the WERNet and then transferred to the reconstruction tasks of the jet flame and the turbulent flame. The encoder shows an extraordinary capability of generalization and the accuracy of the reconstructed result, which even outperforms the original WERNet model without weight encoder transfer. Moreover, the voxel set converges in less than 40 epochs in comparison to 80 epochs in the original WERNet model.

\section{Discussions}
In this investigation, we proposed a brand-new architecture of neural network named WERNet for computed tomography applications with a limited number of views. The WERNet algorithm is a unique algorithm that deposits the sought quantity explicitly in the neural network, and the sought light field/ flame intensity distribution is obtained as the training process is done without interpreting the implicit parameters in the network.  

Essentially, the WERNet algorithm can work well because the limited view tomography is not an overfitting problem and there is much more training data (each ray/pixel can be considered as one piece of training data) than the parameters in the neural network. Therefore, the values of the parameters are well constrained by the training data. Although the weight encoder works as a black box, the $30\times140\times30$ flame intensity data is obtained with a clear physical meaning. As a result, gradient normalization functions in the WERNet model because of the hybridization of two distinctive types of model parameters. Finally, against the intuition, we are not interested in the inputs (impacting voxel sequence) and the outputs (resultant pixel values) from the model comparing with end-to-end or other supervised models.    

Furthermore, the WERNet model can potentially serve in complex realities that only the ray can be traced, but the voxel weight can hardly be obtained. This occurs in the conditions of non-uniform and translucent medium, which deducts the light intensity along the path, and the attenuation of the translucent signal source is influenced by the voxel intensities along the light path, etc. 

Finally, the WERNet model is a semi-case-wise model, as the case-wise refers to the actual flame intensity distribution and cannot be transferred to other applications, and the non-case-wise part refers to the weight encoders that describe the physics of the imaging system. Therefore, this WERNet model is much more general compared with end-to-end algorithms. The hybrid parameter scheme is also valuable in other applications when end-to-end algorithms fail to perform satisfyingly. This feature of the WERNet model potentially can also be applied for other theoretical or complex equation solving cases, such as predicting Navier-Stokes characteristics in fluid dynamics or stress analysis in solid mechanics both for experiments and numerical applications. We expect the formation of the WERNet algorithms can be applied to a boarder range of engineering and fundamental science problems besides the limited-view computer tomography shown in this investigation.

\bibliographystyle{unsrt}  
\bibliography{references}

\end{document}